\documentclass[12pt]{article}
\usepackage[british]{babel}
\def\feynVersion{n}

\def\AnswerYes{y}
\ifx\feynVersion\AnswerYes
   \usepackage{feynmp}                    %%% for Feynman diagrams; a few
\fi

\usepackage[dvips]{color}                 %%% for colours in the graphs
\usepackage{graphicx}                     %%% for pictures and figures
\usepackage{amssymb}
\usepackage{amsmath}
\usepackage{xspace}                       %%% to get the correct spacing of
\newcommand{\3}{\ss}
\newcommand{\non}{\nonumber}
\newcommand{\hq}{\hspace{0.5em}}

\newcommand{\ii}{\mathrm{i}}
\newcommand{\dd}{\mathrm{d}}

\newcommand{\deint}[2]{\dd^{#1}\! #2\;}

\newcommand{\MeV}{\ensuremath{\mathrm{MeV}}}
\newcommand{\fm}{\ensuremath{\mathrm{fm}}}

 \newcommand{\calD}{\mathcal{D}}
\newcommand{\calH}{\mathcal{H}} \newcommand{\calK}{\mathcal{K}}
\newcommand{\calL}{\mathcal{L}} 
 \newcommand{\calO}{\mathcal{O}}
 
\newcommand{\calZ}{\mathcal{Z}}

\newcommand{\mytitle}[1]{\begin{center}\LARGE{\textbf{#1}}\end{center}}
\newcommand{\myauthor}[1]{\textbf{#1}}
\newcommand{\myaddress}[1]{\textit{#1}}
\newcommand{\mypreprint}[1]{\begin{flushright}#1\end{flushright}}

\begin{document}

\begin{titlepage}
  \setcounter{page}{0}

  \mypreprint{
    }

  %\vspace*{0.5cm}
  \vspace*{0.1cm}

  \mytitle{$Nd\rightarrow{^3H}\gamma$ with Effective Field Theory}

  \vspace*{0.5cm}

\begin{center}
  \myauthor{H. Sadeghi}\footnote{Email:
    Hsadegh@chamran.ut.ac.ir} and \myauthor{S. Bayegan}\footnote{Email:
    Bayegan@Khayam.ut.ac.ir}\\[2ex]

  \vspace*{0.5cm}

  \myaddress{Department of Physics, University of Tehran, P.O.Box 14395-547, Tehran, Iran.}
  \\[2ex]

  \vspace*{0.2cm}

\end{center}

\vspace*{0.5cm}

\begin{abstract}
 The cross section of neutron-deuteron radiative capture
  $n d\rightarrow {^3H}\gamma$ is calculated at
energies relevant to Big-Bang nucleosynthesis ( $ 20 \leq E \leq 200
$ KeV ) with pionless Effective Field Theory. At these energies,
magnetic transition M1 gives the dominant contribution. The M1
amplitude is calculated up to next-to-next-to leading order(N$^2$LO)
with insertion of three-body force.
 Results are  in good agreement within few percent theoretical uncertainty
 in comparison with available calculated data below E=200 KeV.
\end{abstract}
\vskip 1.0cm
\noindent
\par\leavevmode\hbox {\it PACS:\ } 26.35.+c, 21.30.Fe, 25.40.Lw, 11.80.Jy, 27.10.+h

\begin{tabular}{rl}
  Keywords: &\begin{minipage}[t]{11cm}
                    effective field theory, three-body system, three-body force,
                    Faddeev equation, radiative capture.
                    \end{minipage}
\end{tabular}

\vskip 1.0cm

\end{titlepage}

\setcounter{footnote}{0}

\newpage

%%%%%%%%%%%%%%%%%%%%%%%%%%%%%%%%%%%%%%%%%%%%%%%%%%%%%%%%%%%%%%%%%%%%%%%%%%%%%%%
%%%%%%%%%%%%%%%%%%%%%%%%%%%%%%%%%%%%%%%%%%%%%%%%%%%%%%%%%%%%%%%%%%%%%%%%%%%%%%%
%%%%%%%%%%%%%%%%%%%%%%%%%%%%%%%%%%%%%%%%%%%%%%%%%%%%%%%%%%%%%%%%%%%%%%%%%%%%%%%
% Main Body
%

%%%%%%%%%%%%%%%%%%%%%%%%%%%%%%%%%%%%%%%%%%%%%%%%%%%%%%%%%%%%%%%%%%%%
\section{Introduction}
\setcounter{equation}{0}
\label{sec:introduction}

 Very low-energy radiative capture and weak capture reactions
involving few-nucleon systems have considerable astrophysical
relevance for studies of stellar structure evolution and big- bang
nucleosynthesis. At these energies pionless Effective field
theory(EFT) is an important tool for computing physical quantities.

 Much of the strength of EFT lies in the fact that it
 can be applied without off shell ambiguities to systems with
 more nucleons. In past years, nuclear EFT has been applied to
 two-, three-, and four-nucleon systems[1-9] and
recently developed pionless EFT is particularly suited to high-order
and precision calculation. An example of a precise calculation  is
the reaction $np\rightarrow \gamma  d$, which is relevant to
big-bang nucleosynthesis(BBN), and  the cross section for this
process was computed to $1\%$ error for center of mass energies
$E\lesssim 1$ MeV~\cite{Rupak99}.

 On the other hand, The three-body EFT
 calculations have been so far confined to nucleon-deuteron system.
 For example nucleon-deuteron scattering in all channels expect the
 $S_{1/2}$-wave can be calculated to high-orders using only two-nucleon
 input  and the triton binding energy is found to be
 ${B_3}^{(EFT)} = 8.35 $MeV in next to leading order close to the
 experimental  ${B_3}^{(exp)} = 8.5 $MeV~\cite{4stooges}.
Pionless EFT has been recently applied to the four-body system with
contact interactions and large scattering length at leading order by
Platter~\cite{Platter}. The binding energies of the ${^4}$He
tetramer and the alpha-particle have been calculated and it is in a
good agreement with experimental value.

 The radiative capture of neutrons and protons
by deuterons and the inverse reactions, the photodisintegration of
$^3$H and $^3$He, have been investigated experimentally and
theoretically over the last decades with some interest. In an
experiment performed in recent years at TUNL~\cite{schmid1,schmid2}
the total cross section and vector and tensor analyzing power of the
$pd\rightarrow {^3He}\gamma $  process were measured at the center
of mass energies below 55 KeV.

 The theory of the $nd\rightarrow{^3H}\gamma $ capture reaction has a
long  history. The $nd$ doublet or quartet state  M1 transition
calculated in Impulse Approximation, and  explanation of smallness
of its cross section when compared to the $np\rightarrow d\gamma$
reaction, $ \sigma_T=334.5\pm0.5$ mb, were pointed out by
Schiff~\cite{schiff}. Later, Phillips~\cite{phillips} emphasized the
importance of initial state interactions and two-body currents to
capture reaction in the three-body model calculation, by considering
a central, separable interaction. In recent years, a series of
calculations of increasing sophistication with regard to the
description of both the initial and final state wave functions and
two-body current model were carried out~\cite{torre}. These efforts
culminated in the 1990 Friar et al.~\cite{friar} calculation of the
$nd\rightarrow {^3H}\gamma $ total cross section, quartet capture
fraction, and photon polarization, based on converged bound and
continuum state Faddeev wave functions, corresponding to a variety
of realistic Hamiltonian models with two- and three-nucleon
interactions, and a nuclear electromagnetic current operator,
including the long-range two-body components associated with pion
exchange and the virtual excitation of intermediate $\Delta$
resonances.

 For very low energy the $p$-$d$ and $n$-$d$ radiative capture,
a magnetic dipole(M1)transition is a dominant contribution,which was
studied in plane wave(Born)approximation by Friar et
al.~\cite{friar}. In these investigations the authors employed their
configuration-space Faddeev calculations of the helium wave
function, with inclusion of three-body forces and pion exchange
currents. Various trends,e.g., the correlation between cross
sections and helium binding energies, and their potential dependence
were pointed out.  More recently a rather detailed investigation of
such processes has been performed by Viviani et~al.~\cite{Viviani}.
Their calculations employed the quite accurate three-nucleon bound-
and continuum states obtained in the variational pair-correlated
hyperspherical method, developed, tested and applied over years by
this group.

 We calculate very low energy cross section of radiative capture of neutrons
by deuterons $n d\rightarrow \gamma {^3H}$ at energies $ 20 \leq E
\leq 200 $ KeV, relevant to Big-Bang nucleosynthesis, with pionless
EFT. At these energies, magnetic transition M1 gives the dominant
contribution. The M1 amplitude is calculated up to next-to-next-to
leading order(N$^2$LO) with insertion of three-body force. Results
show good agreement
 in comparison to ENDF~\cite{ENDF} below E=200 KeV.

The organization of the paper is as follows: We first describe the
relevant Lagrangian and scattering in doublet S-wave channel  in
Section~\ref{section:lagrangian}. This section essentially
introduced to define various parameters that enter in the
expression for the cross section. The calculation of
 the total cross section is presented in sections~\ref{section:M1}. We
tabulate the calculated cross sections
 for some energies relevant for BBN, discuss the theoretical errors, and
 compare our results with the corresponding values from the on-line ENDF/B-VI
database~\cite{ENDF} in Section~\ref{section:results}. Summary and
 conclusions follow in Section~\ref{section:conclusion}.

\section{$^2S_{1/2}$ neutron-deuteron scattering(triton channel)}
\setcounter{equation}{0} \label{section:lagrangian}

The $^2\mathrm{S}_{\frac{1}{2}}$ channel  to which $^3$He and $^3$H
belong  is qualitatively different from the other three-nucleon
channels. Consequently, $^2S_{1/2}$ describes the preferred mode for
$nd\rightarrow{^3H}\gamma$ and $pd\rightarrow{^3He}\gamma$
processes.  This difference can be traced back to the effect of the
exclusion principle and the angular momentum repulsion barrier. In
all the other channels, it is either the Pauli principle or an
angular momentum barrier(or both)which forbids the three-particle to
occupy the same point in space. As a consequence, the kernel
describing the interaction among the three-nucleon,unlike in the
bosonic case,is repulsive in these channels.  The zero mode of the
bosonic case, describes a bound state since it is a solution of the
homogeneous version of the Faddeev equation. As such, it is not
expected to appear in the case of repulsive kernels and, in fact, it
does not. For the $^3$He or $^3$H channel however, the kernel is
attractive and, as we will see below, closely related to the one in
the bosonic case~\cite{4stooges}.

 Let us first discuss the integral equation describing
nucleon-deuteron scattering. We start with the three-nucleon
lagrangian which is given by~\cite{4stooges}:
\begin{equation}\label{triton_deuteron_lag}
\begin{split}
  {\calL}&=N^\dagger\left(\ii\partial_0+\frac{\nabla^2}{2M}\right)N
  +d_s^{A\dagger}\left(-\ii\partial_0-\frac{\nabla^2}{4M}+\Delta_s\right)d_s^A
  +d_t^{i\dagger}\left(-\ii\partial_0-\frac{\nabla^2}{4M}+\Delta_t\right)
  d_t^i\\
  &+t^\dagger\left(\ii\partial_0+\frac{\nabla^2}{6M}+\frac{\gamma^2}{M}+
    \Omega\right)t -g_s\left( d_s^{A \dagger} (N^T P^A N) +\text{H.c.}
  \right) -g_t\left( d_t^{i \dagger} (N^T P^i N) +\text{H.c.}
  \right)\\
  &-\omega_s \left( t^\dagger (\tau^A N) d_s^{A} +\text{H.c.} \right)
  -\omega_t \left( t^\dagger (\sigma^iN) d_t^{i} +\text{H.c.} \right)
  +\dots\;\;,
\end{split}
\end{equation}
where $N$ is the nucleon iso-doublet and the auxiliary fields $t$,
$d_s^A$ and $d_t^i$ carry the quantum numbers of the $^3$He-$^3$H
spin and isospin doublet, $^1S_0$ di-nucleon and the deuteron,
respectively.  The projectors $P^i$ and $P^A$ are defined by:
\begin{eqnarray}\nonumber
  P^i&=&\frac{1}{\sqrt{8}}\tau_2 \sigma_2\sigma^i\\
  P^A&=&\frac{1}{\sqrt{8}}\sigma_2 \tau_2\tau^A \;\;,
\end{eqnarray}
where $A=1,2,3$ and $i=1,2,3$ are iso-triplet and vector indices and
$\tau^A$ ($\sigma^i$) are isospin (spin) Pauli matrices.

\begin{figure}[!htb]
\begin{center}
\includegraphics*[width=.7\textwidth,angle=0,clip=true]{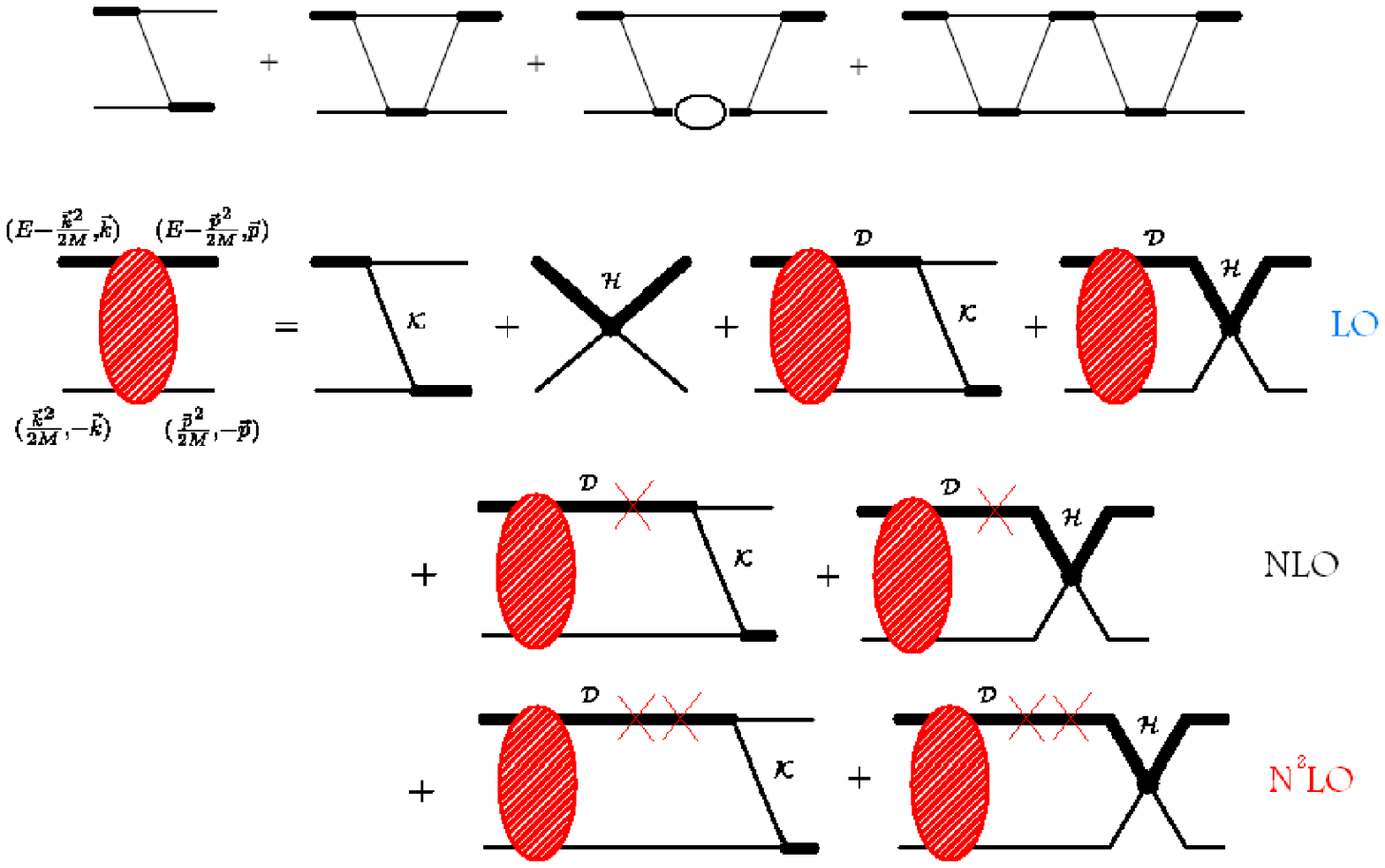}
\caption{The Faddeev equation for $Nd$-scattering up to N$^2$LO.
Thick solid  line: propagator of the two intermediate auxiliary
fields $d_s$ and $d_t$,  denoted by $\calD$, see
(\ref{eq:totaltwoparticlepropagator}); $\calK$:  propagator of the
exchanged nucleon;  $\calH$: three-body force and the cross denotes
insertion of deuteron kinetic energy operator.}
\label{fig:faddeeveq}
\end{center}
\end{figure}

 One can write the Faddeev integral equation in the kinematics defined by the
two cluster-configurations exist in the three-nucleon system.We
follow the notation  suggested by Grie\ss hammer
in~\cite{griesshammer}. The $Nd_t$-cluster with total spin
$S=\frac{3}{2}$ or $S=\frac{1}{2}$, depending on whether the
deuteron and nucleon spins are parallel or anti-parallel; and the
$Nd_s$-cluster which has total spin $S=\frac{1}{2}$, as $d_s^A$ is
a scalar.  The leading-order three-particle amplitude is
$\calO(Q^{-2})$ (before wave-function renormalisation) and
includes all diagrams built out of the leading two-body
interactions. The resultant Faddeev integral equation is
represented in Fig.~\ref{fig:faddeeveq}.
As the Lagrangian up to N$^2$LO does not mix partial waves or flip
the spin of the auxiliary fields, angular momentum is conserved in
the quartet and doublet channels.

In the doublet channel, the Faddeev equation is two-dimensional in
cluster-configuration space as both $Nd_t$- and
$Nd_s$-configurations contribute~\cite{griesshammer}:
\begin{eqnarray}
  \label{eq:doubletpw}
  \vec{t}_d(E;k,p)&=&2\pi\;
  \left[\calK(E;k,p)\;{1\choose -3}
    +\calH(E;\Lambda)\;{1\choose -1}\right]\\
  &&-\;\frac{1}{\pi}\int\limits_0^\infty\deint{}{q} q^2\;\left[
    \calK(E;q,p)\;\begin{pmatrix}1&-3\\-3&1\end{pmatrix}
    +\calH(E;\Lambda)\;\begin{pmatrix}1&-1\\-1&1\end{pmatrix}
  \right]\;
  \non\\&&
  \hq\hq\hq\hq\hq\hq\hq\hq\hq\hq
  \times\;\calD(E-\frac{q^2}{2M},q)\;\vec{t}_d(E;k,q)\;.
  \non
\end{eqnarray}
 The vector
\begin{equation}
  \label{eq:tvector}
  \vec{t}_d := {t_{d,tt}\choose t_{d,ts}}\;,
\end{equation}
is built out of the two amplitudes which get mixed: $t_{d,tt}$ for
the $Nd_t\to Nd_t$-process, and $t_{d,ts}$ for the $Nd_t\to
Nd_s$-process. Furthermore,
\begin{equation}
  \label{eq:totaltwoparticlepropagator}
  \calD(p_0,p):=\begin{pmatrix}D_t(p_0,p)&0\\0&D_s(p_0,p)\end{pmatrix}\;,
\end{equation}
is the propagator of the two intermediate auxiliary fields.

One included the specific three-body force term in the
doublet-channel Faddeev equation (\ref{eq:doubletpw}) at a given
order n of expansion as suggested by Bedaque et
al.~\cite{3stooges_boson,3stooges_doublet,4stooges,griesshammer}~\emph{if
and only if }that term is needed to cancel cut-off dependences in
the observables which are stronger than cut-off dependence from the
suppressed terms of order $(QR)^{n+1}$ where Q is a typical external
momentum and R is the short distance scale in position space beyond
which the EFT breaks down. This argument suggests that from the
point of view of the EFT, cutoff dependences cancel order by order
in the expansion. One can implement this idea by expanding the
kernel of the integral equation perturbatively, and then iterate it
by inserting it into integral equation. This partial re-summation
arbitrarily includes higher order diagrams and consequently dose not
improve the precision of calculation. The only necessary
re-summation is the one present at LO. The partial re-summation of
range effects is made only for convenience~\cite{3stooges_boson}.

 The integral equation is solved numerically by imposing a cut-off
$\Lambda$. In that case, a unique solution exists in the
$^2S_{1/2}$-channel for each $\Lambda$ and $\calH=0$, but no unique
limit as $\Lambda\to\infty$.  As long-distance phenomena must
however be insensitive to details of the short-distance physics (and
in particular of the regulator chosen), Bedaque et
al.~\cite{3stooges_boson,3stooges_doublet,4stooges,griesshammer}
showed that the system must be stabilized by a three-body force
\begin{equation}
  \label{eq:calH}
   \calH(E;\Lambda)=
   \frac{2}{\Lambda^2}\sum\limits_{n=0}^\infty\;H_{2n}(\Lambda)\;
   \left(\frac{ME+\gamma_t^2}{\Lambda^2}\right)^n
   =\frac{2H_0(\Lambda)}{\Lambda^2}+
   \frac{2H_2(\Lambda)}{\Lambda^4}\;(ME+\gamma_t^2)+\dots \;.
\end{equation}
which absorbs all dependence on the cut-off as $\Lambda\to\infty$.
It is analytical in $E$ and can be obtained from a three-body
Lagrangian, employing a three-nucleon auxiliary field analogous to
the treatment of the two-nucleon channels~\cite{4stooges}.

$H_2$ is dimension-less but depends on the cut-off $\Lambda$ in a
non-trivial way, as a renormali\-sation group analysis reveals:
Instead of approaching a fixed-point as $\Lambda\to\infty$, it shows
an oscillatory behavior known as ``limit
cycle''~\cite{3stooges_boson}.

As one needs a three-body force at LO, $H_0\sim Q^{-2}$, all
three-body forces obtained by expanding $\calH$ in powers of $E$ are
also enhanced, with the interactions proportional to $H_2$ entering
at N$^2$LO~\cite{4stooges,griesshammer}. The power-counting for the
three-body forces is hence
\begin{equation}
  H_0(\Lambda)\sim Q^{-2}\;\;,\;\;H_2(\Lambda)\sim Q^{-2}\;\;,
  \;\;H_{2n}(\Lambda)\sim Q^{-2} .
\end{equation}

 The scattering phase-shift of the
S-wave in the quartet and doublet channel is related to the
renormalised on-shell amplitudes by
\begin{equation}
 T_q=\calZ_t\;t_q=
  \frac{3\pi}{M}\;\frac{1}{k \cot\delta_q-\ii k}\;\;,\;\;
  T_{d,xy}=
  \frac{3\pi}{M}\;\frac{1}{k \cot\delta_{d,xy}-\ii k}\;\; ,
  \label{eq:phaseshifts}
\end{equation}

where $x,y=s,t$ label the matrix entries in cluster-configuration
space, and
\begin{equation}
  \vec{T}_{d}=\calZ \vec{t}_{d} \;\;\mbox{ with }
  \;\;\calZ:=\begin{pmatrix}\calZ_t&0\\0&\sqrt{\calZ_t\calZ_s}\end{pmatrix}\;,
\end{equation}
is the renormalised doublet-amplitude and its wave-function
renormalisation. In the doublet channel, the only observable process
is nucleon-deuteron scattering, $Nd_t\to Nd_t$, i.e.~$x=y=t$.

Nucleon-deuteron scattering is to N$^2$LO thus completely determined
by four simple observables of $NN$-scattering: the deuteron binding
energy,residue (or effective range),the scattering length and
effective range of the $^1S_{0}$-channel. Only the
$^2S_{1/2}$-channel has further unknowns, namely the strength of the
three-body interaction $H_0$ at LO and NLO, and in addition of $H_2$
at N$^2$LO . They are determined by its measured scattering length
$a_d$~\cite{doublet_sca} and the triton binding energy $B_d$,
respectively:
\begin{equation}
  \label{eq:threebodyexpvalues}
  a_d=(0.65\pm0.04)\;\fm\;\;,\;\;B_d=8.48\;\MeV \;.
\end{equation}

\section{ Neutron-deuteron radiative capture process}
\setcounter{equation}{0} \label{section:M1}

 In the KeV energy region, the processes $nd\to {^3H}\gamma$ and
$pd\to {^3He}\gamma$ play an important role in nuclear astrophysics
and in nuclear physics. In the standard big-bang nucleosynthesis
theory the corresponding reaction rates are necessary to estimate
the ${^3}$He-yield as well as the abundances of other light
elements. The spin structure of the matrix elements of neutron
radiative capture by deuteron is complicated but in very low energy
for this reaction we can introduced three multipole transition that
can be allowed by p-parity and angular momentum conservation i.e.
$J^p={\frac{1}{2}}^+\rightarrow M1$ and
$J^p={\frac{3}{2}}^+\rightarrow M1, E2$. The parameterization of the
corresponding contribution to the matrix element to be build by the
following contributions:

$$i(t^\dagger N)(\vec D\cdot\vec{e^*}\times\vec k)\;,$$
\begin{equation}
(t^\dagger\sigma_a N)(\vec D\times [\vec{ e^*}\times\vec k])_a\;,
\label{eq:as7}
\end{equation}
$$t^\dagger(\vec\sigma\cdot\vec{e^*}~\vec D\cdot\vec k+\vec\sigma\cdot
\vec k~ \vec D\cdot\vec e^*)N\;,$$ where $N$, $t$, $\vec e$, $\vec
D$ and $\vec k$ are the 2-component spinors of initial nucleon
filed, final ${^3H}e$ (or ${^3}H$)field, the 3-vector polarization
of the produced photon, the 3-vector polarization of deuteron and
the unit vector along the 3-momentum of the photons, respectively.
The two structures in Eq.(\ref{eq:as7}) correspond to the M1
transition. At thermal energies the reaction proceeds through S-wave
capture predominantly via magnetic dipole transition, $M^{LSJ}_{i}$,
where L=0, $S$=1/2,3/2 and i=1. To obtain the spin structure, which
corresponds to a definite value of $J$ for the entrance channel, it
is necessary to build special linear combinations of products $\vec
DN$ and $\vec \sigma\times\vec DN$, with
$J^{P}=\displaystyle\frac{1}{2}^+$ or
$J^{P}=\displaystyle\frac{3}{2}^+$:
$$\vec {\phi}_{1/2}=(i\vec D+\vec\sigma\times\vec D)N~\mbox{and}~(2i\vec
D-\vec\sigma\times\vec D)N\;.$$
 For both possible magnetic dipole transitions with
 $ J^{P}=\displaystyle\frac{1}{2}^+$ (amplitude $g_1$) and
  $J^{P}=\displaystyle\frac{3}{2}^+$ (amplitude $g_3$) we can write:
$$g_1:~~t^\dagger(i\vec D\cdot\vec{e^*}\times\vec k+\vec\sigma\times\vec
D\cdot\vec{e^*}\times\vec k)N,$$
\begin{equation}
g_3:~~t^\dagger(i\vec D\cdot\vec{e^*}\times\vec k+\vec
\sigma\times\vec D\cdot\vec{e^*}\times\vec k)N\;. \label{eq:as8}
\end{equation}

 The electric transition $E^{LSJ}_i$
for energies of less than 60 KeV dose not contribute to the total
cross section. Therefore $E^{0 (3/2) (3/2)}_2$ transition will not
be considered in energies relevant to BBN calculation. The M1
amplitude receives contributions from the magnetic moments of the
nucleon and dibaryon operators coupling to the magnetic field, which
are described by the lagrange density involving fields:
\begin{equation}\label{eq:M1}
  \mathcal{L}_B=\frac{e}{2M_N}N^\dagger(k_0+k_1 \tau^3){\sigma.B}
  +e\frac{L_1}{M_N\sqrt{r^{({^1s}_0)}r^{({^3S}_1)}}}{{d_t}^j}^\dagger{{d_s}}_3
  B_j+H.C\;.
\end{equation}
where the unknown coefficient $L_1$, which contributes at order
$Q$ must either be predicted from QCD or determined experimentally
in order to have model-independent predictive power.
\begin{figure}[!htb]
\begin{center}
 \includegraphics*[width=.7\textwidth]{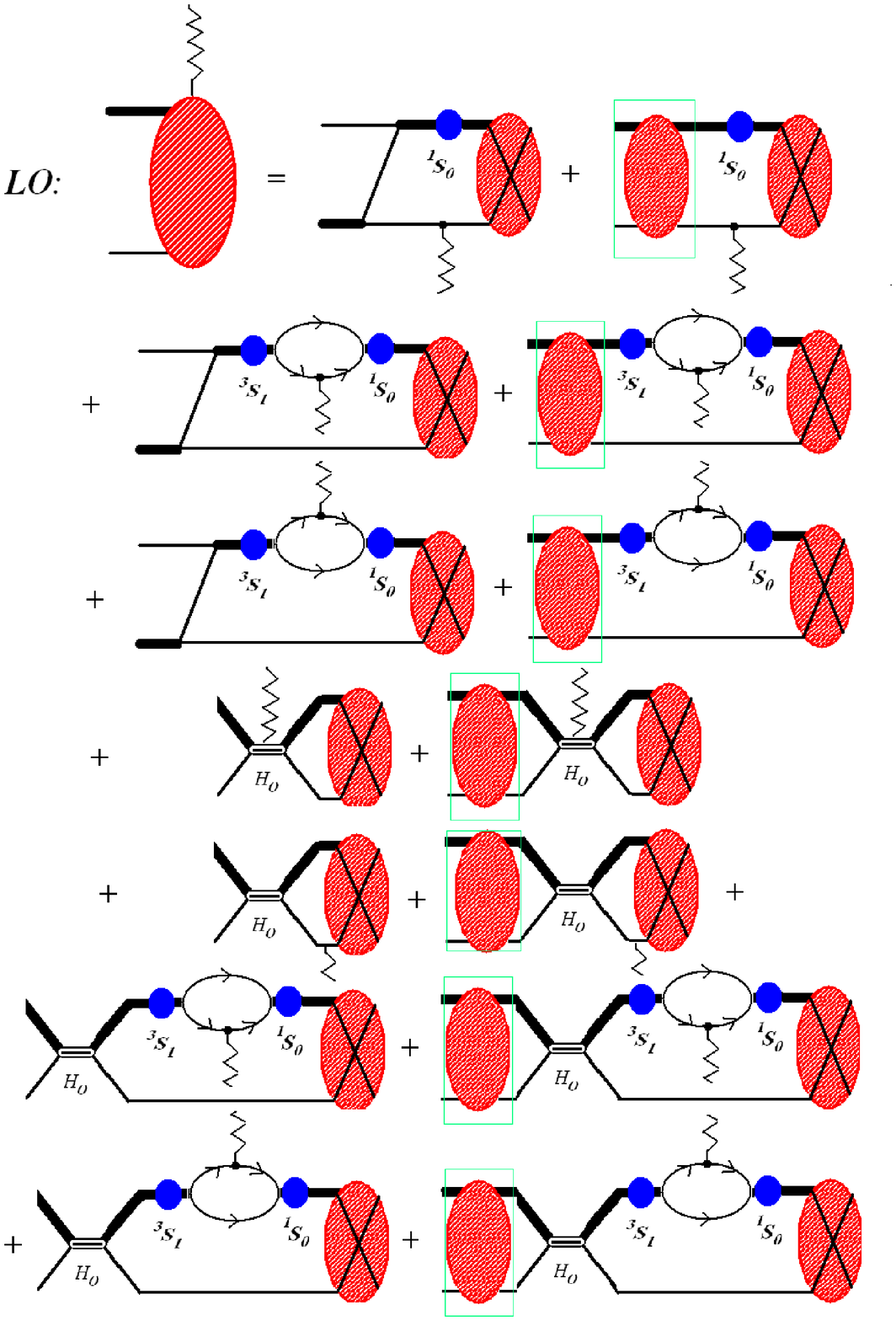}
 \caption{The Faddeev equation for $Nd$-radiative capture  up to LO. Boxes indicate insertion of
  $Nd$-scattering amplitude up to LO from Fig.~\ref{fig:faddeeveq}(only up to first line of perturbative expansion of
  the Faddeev equation).  Cross shows
  wave function renormalization in each step when triton is made and three-body
  interactions are shown with strength $H_0(\Lambda)$. Wavy line shows photon and small circles show magnetic photon
  interaction. The photon is minimally coupled.
  Remaining notation as in Fig.~\ref{fig:faddeeveq}.} \label{fig1}
\end{center}
\end{figure}

\begin{figure}[!htb]
\begin{center}
  \includegraphics[width=0.7\linewidth,angle=0,clip=true]{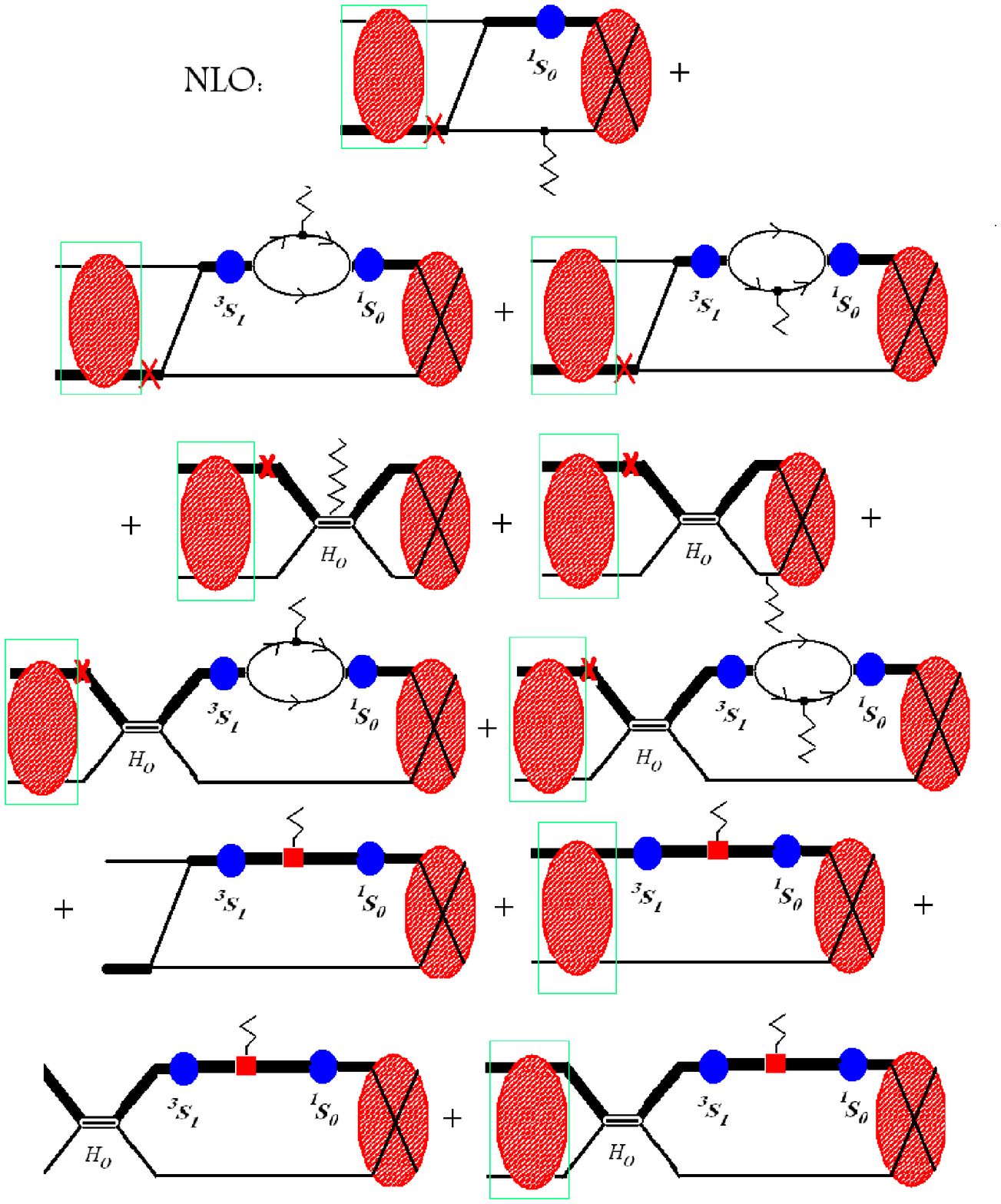}
\end{center}
\vspace*{-0pt} \caption{ The NLO  contribution can then be obtained
by perturbing around the LO solution with the one  deuteron kinetic
energy operator insertion and last lines show NLO order of photon
interaction with the lagrangian Eq.(\ref{eq:M1}) with the $L_1$
vertices. Boxes indicate insertion of
  $Nd$-scattering amplitude for NLO from Fig.~\ref{fig:faddeeveq}(only second line of perturbative expansion of
  the Faddeev equation). Remaining
  notation as in Fig.~\ref{fig1}.} \label{fig2}
\end{figure}

%%%%%%%%%%%%%%%%%%%%%%%%%%%%%%%%%%%%%%%%%%%%%%%%%%%%%%%%%%
\begin{figure}[!htb]
\begin{center}
  \includegraphics[width=0.7\linewidth,angle=0,clip=true]{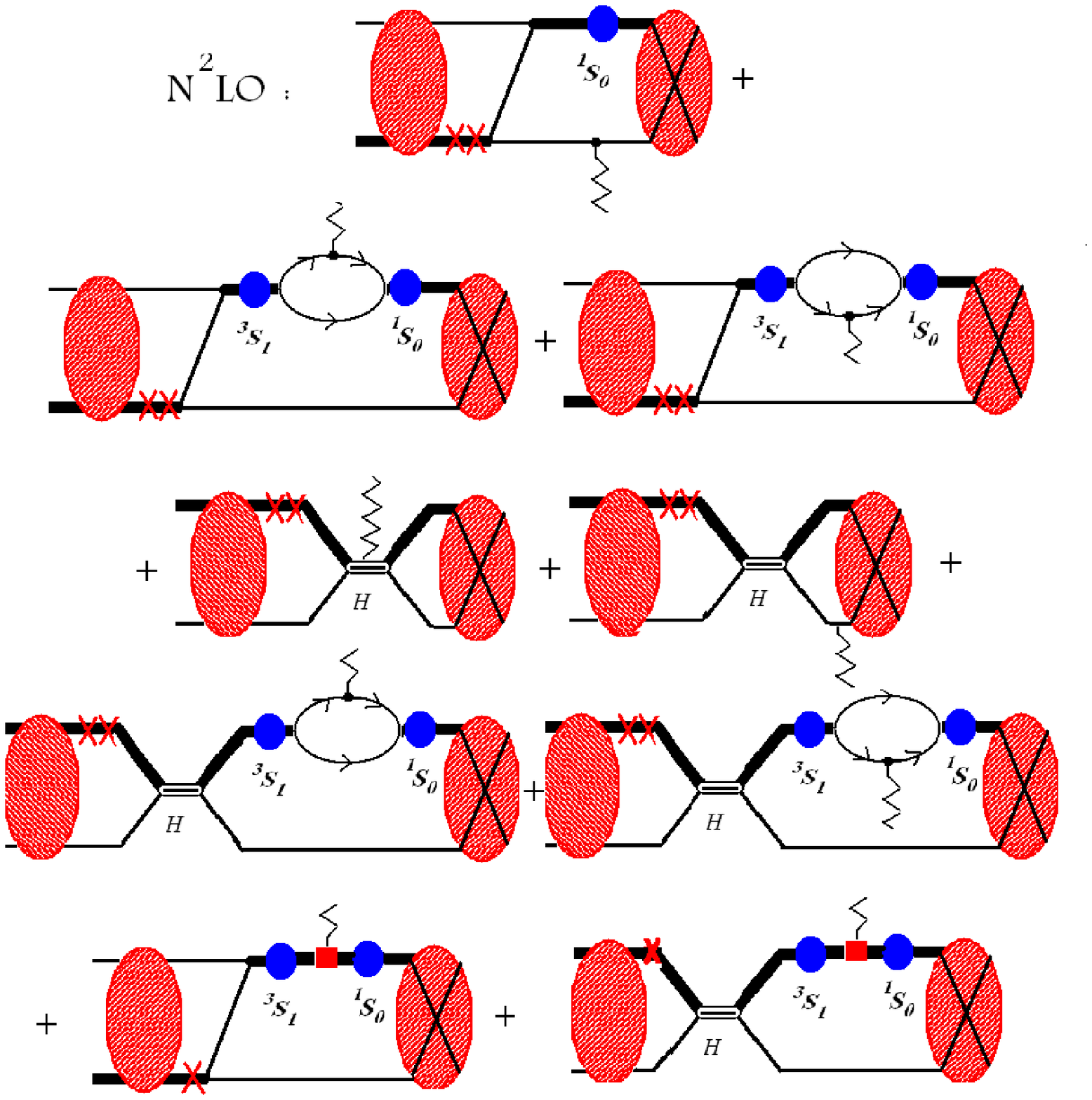}
\end{center}
\vspace*{-0pt} \caption{The N$^2$LO  contribution can then be
obtained by perturbing around the LO solution with the twice
deuteron kinetic energy operator insertion. Boxes indicate insertion
of $Nd$-scattering amplitude to N$^2$LO from
Fig.~\ref{fig:faddeeveq}. Remaining
  notation as in Fig.~\ref{fig1} and Fig.~\ref{fig2}. }
\label{fig3}
\end{figure}
%%%%%%%%%%%%%%%%%%%%%%%%%%%%%%%%%%%%%%%%%%%%%%%%%%%%%%%%%%%%%%
%%%%%%%%%%%%%%%%%%%%%%%%%%%%%%%%%%%%%%%%%%%%%%%%%%%%%%%%%%
\begin{figure}[!htb]
\begin{center}
  \includegraphics[width=0.4\linewidth,angle=0,clip=true]{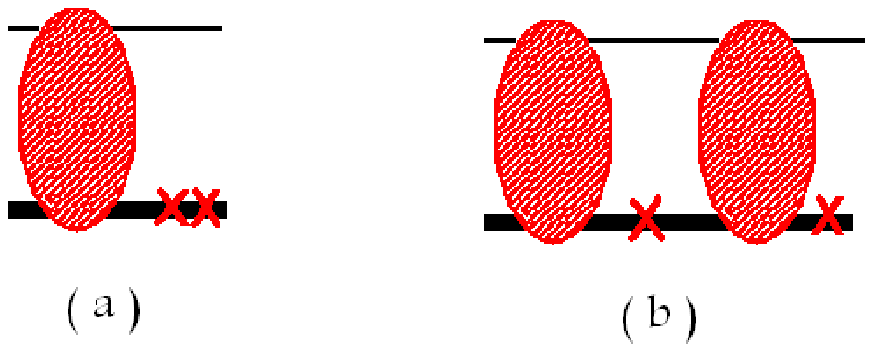}
\end{center}
\vspace*{-0pt} \caption{The two different contribution of N$^2$LO
with twice deuteron kinetic energy operator insertion.} \label{fig4}
\end{figure}
%%%%%%%%%%%%%%%%%%%%%%%%%%%%%%%%%%%%%%%%%%%%%%%%%%%%%%%%%%
The radiative capture cross section $nd\rightarrow {^3H}\gamma$ at
very low energy is given by

\begin{equation}\label{crossection}
  \sigma=\frac{2}{9}\frac{\alpha}{v_{rel}}\frac{p^3}{4M^2_N}\sum_{iLSJ}
  [{|\widetilde{\chi}^{LSJ}_i|}^2]\;,
\end{equation}
where
\begin{equation}\label{redefine}
  \widetilde{\chi}^{LSJ}_i=\frac{\sqrt{6\pi}}{p\mu_N} \sqrt{4\pi}
  {\chi^{LSJ}_i}\;,
\end{equation}
with $\chi$ stands for either the electric or magnetic transition
and $\mu_N$ is in nuclear magneton and p is momentum of the incident
neutron in the center of mass. There is an infinite number of
diagrams contributing at leading order for M1 amplitude of the
radiative capture cross section $nd\rightarrow {^3H}\gamma$, as
shown in Fig.~\ref{fig1}. We follow the same procedure as
~\cite{4stooges}: First, expand the kernel of the integral equation
perturbatively in the same way of Fig.~\ref{fig:faddeeveq}, and then
 iterate it by inserting it into the integral equation. We
use the deuteron propagator in the integral equation, which is then
solved numerically. The re-summation is necessary since according to
the power counting  all these diagrams contribute equally to the
final amplitude.  The NLO and N$^2$LO contributions can then be
obtained by perturbing around the LO solution with the one and twice
deuteron kinetic energy operator insertion, respectively, as shown
diagrammatically in Fig.~\ref{fig2} and Fig.~\ref{fig3}. For N$^2$LO
contribution due to very cumbersome numerical calculation, one must
substitute diagram of Fig.~\ref{fig4}(a) by Fig.~\ref{fig4}(b) in
diagrams of Fig.~\ref{fig3}, particularly when one needs to compute
the full off-shell LO amplitude before inserting the deuteron
kinetic energy terms.

We now turn to  the amplitude to be used in the $M1$ calculation.
The amplitude $\vec{t}_d $ after properly iterated  will be folded
to electromagnetic interaction  $\mathcal{K}_{EM}$ order by order
and integrated on momentum in each order. Finally  the wave function
renormalization in each order will be done when triton is made. we
introduce for $\mathcal{K}_{EM}$ :

\begin{eqnarray}
\mathcal{K}_{EM} & = & {1\over\sqrt{1-\gamma r^{({^3}S_1)} }} \
{1\over -{1\over a^{({^1}S_0)}} + {1\over 2} r^{({^1}S_0)} |{\bf
q}|^2 - i|{\bf q}|}
\nonumber\\
& & \left[\ \kappa_1 \ {\gamma^2\over |{\bf q}|^2 + \gamma^2} \left(
\gamma - {1\over a^{({^1}S_0 )}} + {1\over 2} r^{({^1}S_0)} |{\bf
q}|^2 \right) \ +\ L_1\  {\gamma^2\over 2} \ \right] \; .
\label{eq:xmonet}
\end{eqnarray}

\begin{figure}[!htb]
\begin{center}
   \includegraphics*[width=.2\textwidth,angle=270,clip=true]{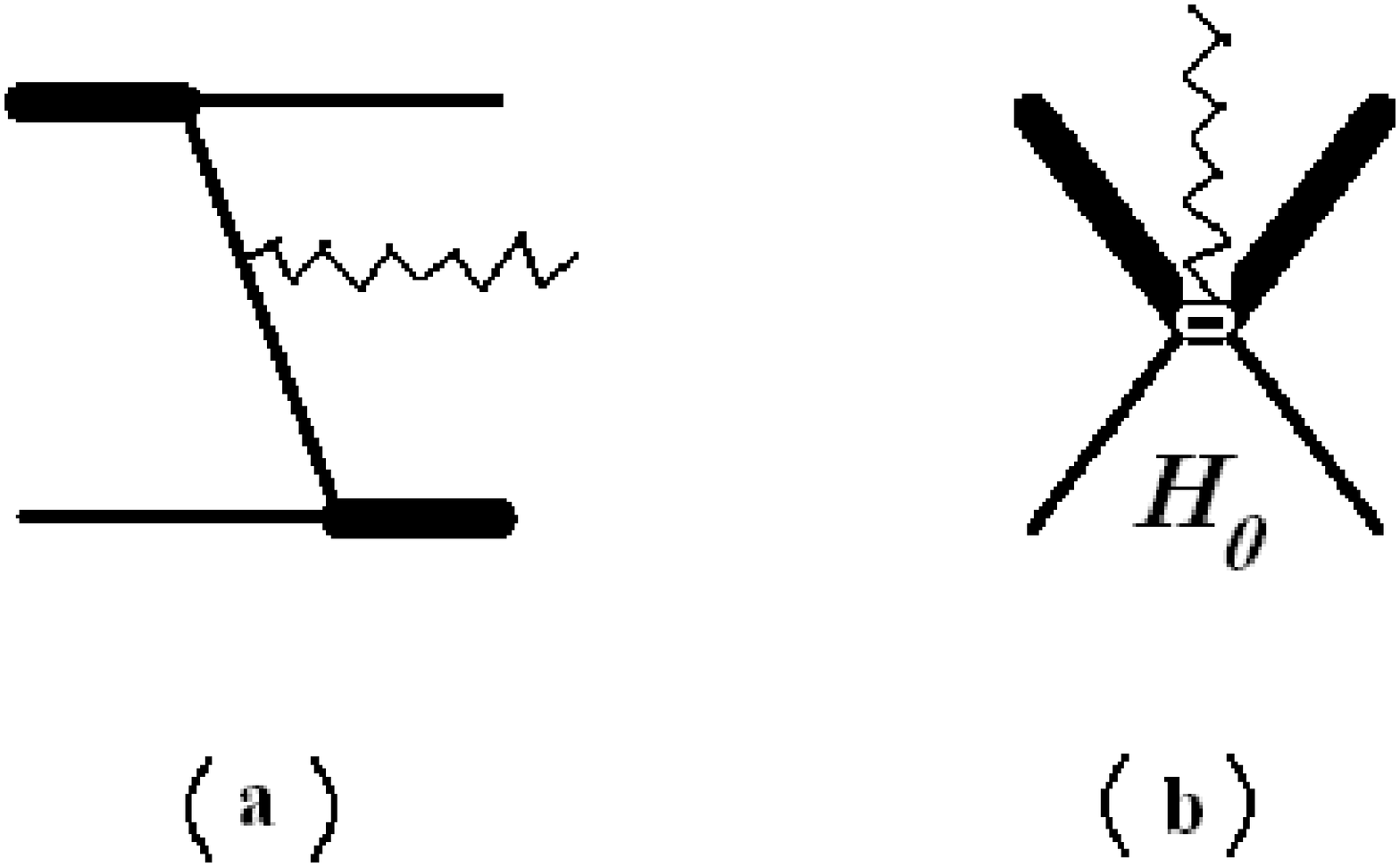}
\caption{Diagram (a) shows photon interaction with exchanged nucleon
and diagram (b) shows interaction of photon with $H_0$ vertex.}
\label{fig5}
\end{center}
\end{figure}

We have not considered the two  following  possible diagrams in our
calculation.  Photo-interaction directly with exchanged nucleon is
shown in Fig.~\ref{fig5}(a). For this interaction, we have $p.A$ and
in very low energy relevant to BBN,  $p \leq 200$ KeV ,this
particular contribution (Fig.~\ref{fig5}(a)) which could be appeared
in Figures (2,3,4) has been neglected, because we can estimate it's
size from a naive power counting.  In the integration of photon
coupling to exchanged-nucleon vertices,  we use $q \sim Q$ and $q
\sim p$, where we have the $Q/M^2y^2$ the wave function
renormalization, $M/Q^2$ nucleon propagator and $1/My^2Q$ dibaryon
propagator. Consequently, it's size of contribution has been
estimated to be $< 1\%$. The other direct interaction is with
three-body vertices $\mathcal{H}$, up to order of our calculation
only $H_0$, $H_2$ are considered. The direct interaction in order of
$H_0$, for zero energy range, will be dealt with elsewhere. This
diagram is shown in Fig.~\ref{fig5}(b). Contribution of this diagram
for the energy range of our calculation is ignored, because in very
low energy for $p\sim q\sim Q$ error for neglecting the contribution
of this term is less than 1$\%$. Contribution of diagrams with $H_2$
vertices generally appears at N$^2$LO. At N$^2$LO one can insert
photon also to $H_2$ vertices. From (\ref{eq:calH}), this
contribution is $(p^2/\Lambda^4)H_2$ and is of the order of
$(k/\Lambda)^2$ or $(Q/\Lambda)^2$ by power counting expansion. In
numerical calculation we define $H_2$ such that it dose not
contribute at zero momentum, naturally for the energy range near
zero momentum, insertion of photon to $H_2$ vertices for momentum
$p\sim 200$ KeV, could be neglected and the error is estimated to be
smaller than the estimation of error for insertion of photon to
$H_0$ vertices.

The solution $\vec{t}_d $ depends on  $\Lambda$ for arbitrary
$H_0,H_2$ which in general are cutoff dependent, too. Since the
low-energy amplitude can not depend on the cutoff, any explicit
dependence on the cutoff has to be canceled by the cutoff dependence
implicit in $H_0,H_2$, etc. We determine the cutoff dependence of
the three-body forces by imposing this condition order by order in
$1/\Lambda$ in the same way of ~\cite{4stooges}. In this process, we
also determine which three-body force appears at every order of the
expansion.

In order to have  cutoff independent amplitude  one  must determine
the three-body forces  by demanding order by order in $Q/\Lambda$

\begin{equation}\label{master_renorm}
  {\mathcal H}(E,\Lambda) + \frac{2}{\pi}\int\limits^\Lambda \dd q\; q^2
  \mathcal{D} (E-\frac{q^2}{2M},q)
  \left[ {\mathcal K}(q,p) + {\mathcal H}(E,\Lambda)
  \right]t_\Lambda(q)=\mathrm{const.}\;\;,
\end{equation}
This (possibly energy dependent) constant can without loss of
generality be set to zero by absorbing it into a re-defined three
body force
\begin{equation}
  \mathcal{H}(E,\Lambda)\to\mathcal{H}(E,\Lambda)-
  \frac{\mathrm{const.}}{1+\displaystyle
    \frac{2}{\pi}\int\limits^\Lambda \dd q\; q^2
  \mathcal{D} (E-\frac{q^2}{2M},q)\;t_\Lambda(q)}\;\;.
\end{equation}

\begin{figure}[!htb]
\begin{center}
   \includegraphics*[width=.8\textwidth,angle=0,clip=true]{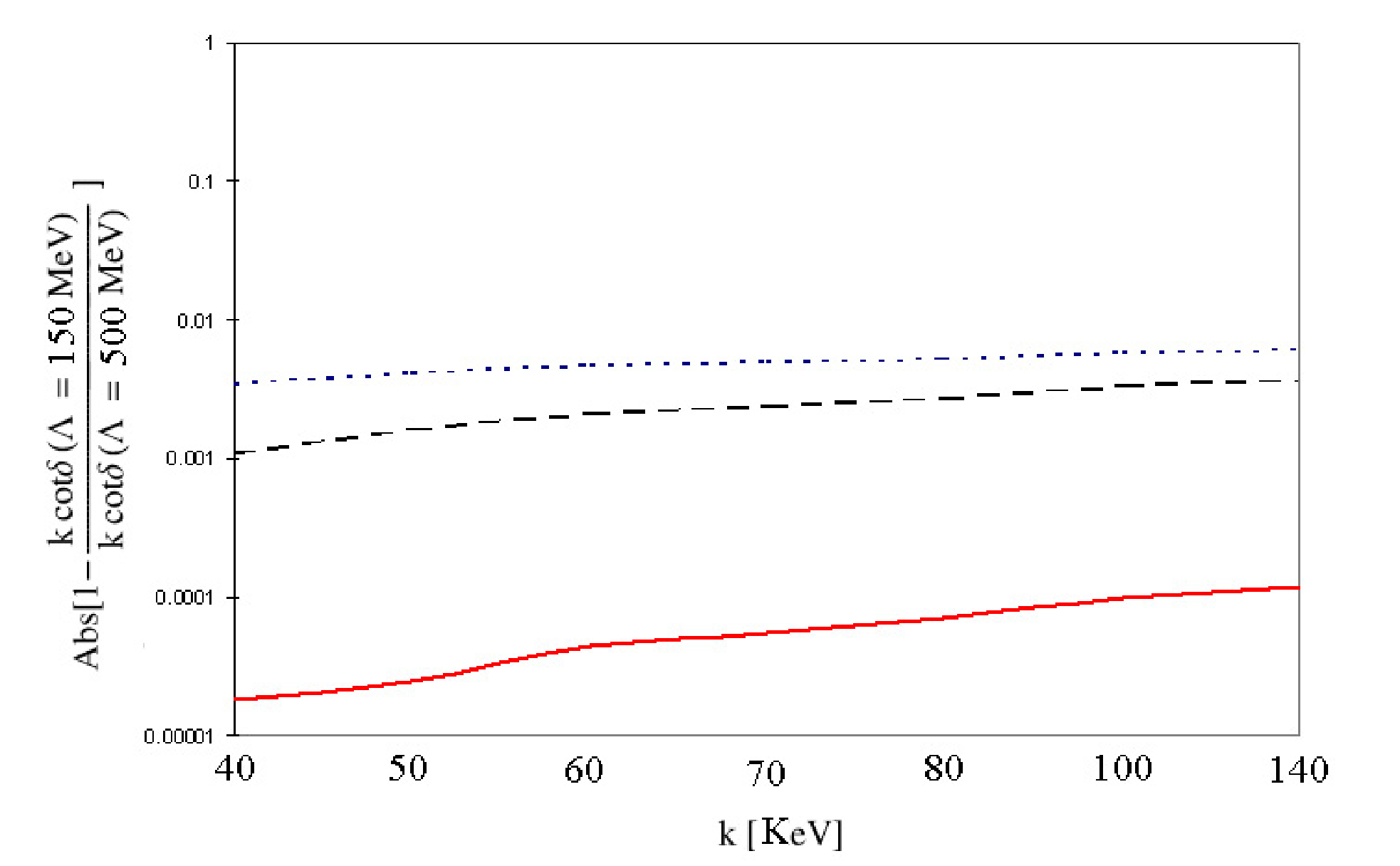}
\caption{Semi-logarithm curve of the cutoff variation of phase
shifts is shown between  $\Lambda=150$ MeV and $\Lambda=500$ MeV as
function of the center-of-mass momentum .  The short dashed, long
dashed and solid line correspond to LO, NLO and N$^2$LO,
respectively.} \label{fig:cutoff}
\end{center}
\end{figure}
Imposing (\ref{master_renorm}) with $\mathrm{const.}=0$ order by
order in the low energy expansion, we can determine which three-body
forces are required at any given order, and how they depend on the
cutoff. As we are interested only in the UV behavior, i.e.~in the
asymptotics, and no IR divergences occur, we can safely neglect the
IR limit of the integral. All calculations were performed with the
cutoff at $400$ MeV. At  N$^2$LO , where  we saw that $H_2$  is
required the cutoff is varied between $150$ and $500$ MeV and we
ignore other sources  of error at low cutoffs, this is a reasonable
estimate of the errors.  For energy range of our calculation, below
the break up reaction, we show cutoff variation of phase shifts
between $\Lambda=150$ MeV and $\Lambda=500$ MeV as function of the
center-of-mass momentum in Fig.~\ref{fig:cutoff}. We can see very
smooth slope and does however change significantly from order to
order because the dominant correction is $~(\gamma/\Lambda)^n$ and
for these energy range at every order these variations are nearly
independent of variation of momentum. We confirm that in our
calculation also in very low energy range the cutoff variation
decreases steadily as we increase the order of the calculation and
it is of the order of $(k/\Lambda)^n, (\gamma/\Lambda)^n$, where $n$
is the order of the calculation and $\Lambda=150$ MeV is the
smallest cutoff used.It is worth mentioning that the errors  due to
increasing momentum, as one would expect due to $(k/\Lambda)^n$ also
appear in our calculation in very low energy but these errors
decrease when the order of calculation is increased up to N$^2$LO.

%%%%%%%%%%%%%%%%%%%%%%%%%%%%%%%%%%%%%%%%%%%%%%%%%%%%%%%%%%%%%%%%%%%%%%%%%%%%%%%
\section{Numerical results for  neutron-deuteron radiative capture}
\label{section:results}
%%%%%%%%%%%%%%%%%%%%%%%%%%%%%%%%%%%%%%%%%%%%%%%%%%%%%%%%%%%%%%%%%%%%%%%%%%%%%%%
%%%%%%%%%%%%%%%%%%%%%%%%%%%%%%%%%%%%%%%%%%%%%%%%%%%%%%%%%%

 We numerically solved the Faddeev integral equation up to N$^2$LO. We used $\hbar
c=197.327\;\MeV\,\fm$, a nucleon mass of $M=938.918\;\MeV$, for the
$NN$ triplet channel a deuteron binding energy (momentum) of
$B=2.225\;\MeV$ ($\gamma_d=45.7066\;\MeV$), a residue of
$Z_d=1.690(3)$, effective range $r_{0t}=2.73\;\fm$, for the $NN$
singlet channel an ${}^1\mathrm{S}_0$ scattering length of
$a_t=-23.714\;\fm$ and $\mu_N=5.050\times10^{-27} JT^{-1}$. We
determine the two-nucleon parameters from the deuteron binding
energy, triplet effective range (defined by an expansion around the
deuteron pole, not at zero momentum), the singlet scattering length,
effective range (defined by expanding at zero momentum), and two
body capture process(obtained with comparison between experimental
data and theoretical results for $nd \rightarrow d \gamma$ process
at zero energy~\cite{Rupak99}). We fix the three-body parameters as
follows: because we defined $H_2$ such that it does not contribute
at zero momentum scattering, one can first determine $H_0$ from the
${}^2\mathrm{S}_\frac{1}{2}$ scattering length
$a_3=(0.65\pm0.04)\;\mathrm{fm}$~\cite{doublet_sca}.
\begin{figure}[!htb]
\begin{center}
  \includegraphics[width=0.5\linewidth,angle=270,clip=true]{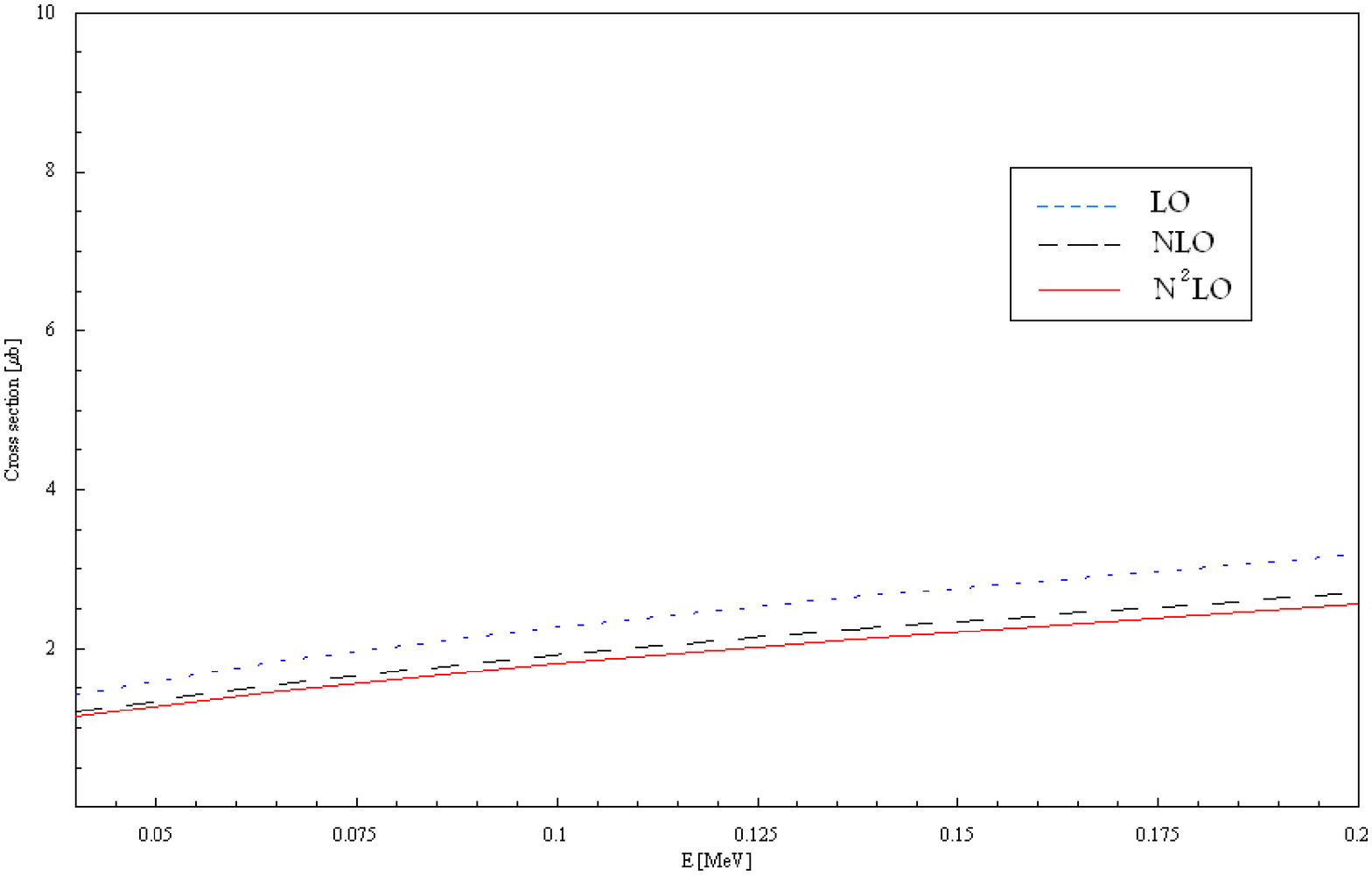}
\end{center}
\vspace*{-0pt} \caption{The cross section for neutron radiative
capture by deuteron as function of the center-of-mass
  kinetic energy $E$ in Mev.  The short dashed, long dashed and solid line correspond
  to the  contribution from M1 capture up to LO, NLO and N$^2$LO, respectively.}
\label{crosssection}
\end{figure}

\begin{table}[!htb]
\caption{Neutron radiative capture by deuteron in micro barn and
estimated error in percentage in comparison with the last column at
every order up to N$^2$LO. Last column shows ENDF results for cross
section~\cite{ENDF}.} \label{tab:a} \vspace{0.25cm}
\begin{center}
\begin{tabular}{c c||c|c|c|c|c|c|c}
\hline &Energy  & $\sigma(\mu b)$ & error($\%$) & $\sigma(\mu b)$ &
error($\%$)&
$\sigma(\mu b)$ & error($\%$)& $ENDF(\mu b)$ \\
&(KeV)  & LO & LO & NLO &  NLO& N$^2$LO &  N$^2$LO&  \\
 \hline \hline

 & 40 & 1.64(6)& 30& 1.31(5)  & 4 & 1.25(0) & 1.5& 1.27(0)\\
 & 50 & 1.72(8)& 24& 1.45(8) &4 & 1.38(3) & 0.7& 1.39(0)\\
 & 60 & 1.90(5)&27 & 1.58(5)  & 5 &1.50(1)& 0.1&1.50(0)\\
 & 70 & 2.02(6)& 25& 1.72(0) & 7 &1.61(6) &0.1 & 1.61(0)\\
 & 80 & 2.11(9)& 22& 1.82(2)  & 6 &1.73 (3)& 0.5& 1.72(0)\\
 & 100 & 2.42(6)& 25& 2.04(4)  &  5&1.93(3) & 0.5& 1.94(0)\\
 & 140 & 2.88(9) & 30& 2.41(8)  & 9& 2.30(3) & 0.1 &2.22(9)\\
\hline
\end{tabular}
\end{center}
\end{table}
 At LO and NLO,
this is the only three-body force entering, but at N$^2$LO, where we
saw that $H_2$ is required, it is determined by the triton binding
energy $B_3=8.48\;\mathrm{MeV}$. We solve integral equation by
expansion  in order of $Q$  and  properly iterating  the kernel.
Then, the resulted $\vec{t}_d $  will be folded to electromagnetic
interaction order by order and  properly integrated on the involving
momentum.
\begin{figure}[!htb]
\begin{center}
  \includegraphics[width=0.7\linewidth,angle=0,clip=true]{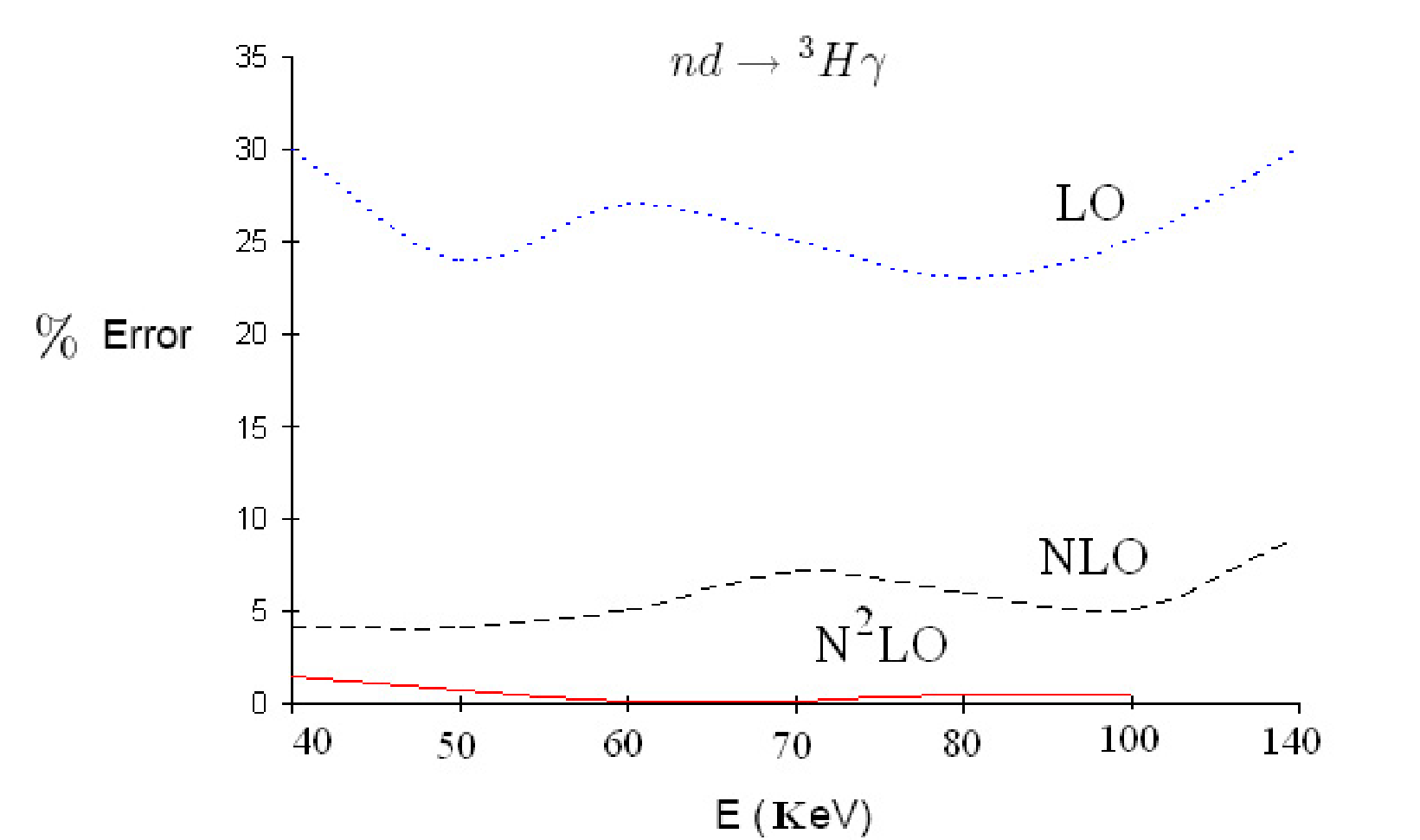}
\end{center}
\vspace*{-0pt} \caption{Estimation of error in percentage is shown
in comparison with ENDF~\cite{ENDF} versus nucleon center-of-mass
  kinetic energy $E$ in KeV.  The short dashed, long dashed and solid line correspond
  to the error up to LO, NLO and N$^2$LO, respectively.}
\label{error}
\end{figure}

For our calculation source of error due to low cutoffs and low
momentums for very low energy calculation is neglected.   The cutoff
variation decreases steadily as we increase the order of the
calculation and is of the order of $(k/\Lambda)^n,
(\gamma/\Lambda)^n$, where $n$ is the order of the calculation. We
used  $\Lambda=150$ MeV, for the smallest cutoff .

 The  cross section  calculation  for neutron radiative capture by deuteron as
function of the center-of-mass energy  at LO, NLO and N$^2$LO is
shown in Fig.~\ref{crosssection}.

 Table {1} shows numerical results
for the EFT $nd\rightarrow {^3H}\gamma$ cross section for various
nucleon center of mass energies E up to N$^2$LO and errors estimate
at every order in comparison with the last column. The corresponding
values for the cross section from the online evaluated nuclear data
file ENDF/B-VI ~\cite{ENDF}  are shown in the last column. The EFT
results for this cross section are presented up to only two
significant digits(the third digit is shown for better comparison
with ENDF data). In Fig.~\ref{error}, we show the error due to
available evaluated data ENDF~\cite{ENDF} in percentage versus
nucleon center-of-mass kinetic energy $E$ in KeV.

%%%%%%%%%%%%%%%%%%%%%%%%%%%%%%%%%%%%%%%%%%%%%%%%%%%%%%%%%%%%%%%%%%%%%%%%%%%%%%%
%%%%%%%%%%%%%%%%%%%%%%%%%%%%%%%%%%%%%%%%%%%%%%%%%%%%%%%%%%%%%%%%%%%%%%%%%%%%%%%
\section{Conclusion}
\label{section:conclusion}
%%%%%%%%%%%%%%%%%%%%%%%%%%%%%%%%%%%%%%%%%%%%%%%%%%%%%%%%%%%%%%%%%%%%%%%%%%%%%%%

We have calculated the cross section of radiative capture process
$nd\rightarrow {^3H}\gamma$. We applied pionless EFT to find
numerical results for the M1 contributions for this capture process
for incident neutron energies relevant for BBN, $0.02\leq E \leq
0.2$ MeV. At these energy our calculation is dominated by S-wave
state and magnetic transition M1 contribution.

The error estimate in the cross section in comparison with evaluated
nuclear data file ENDF~\cite{ENDF} is shown in Fig.~\ref{error} and
Table {1}.  Errors estimate are 20-30 percent at leading order,
below 10 percent up to NLO and by insertion of three-body force at
N$^2$LO, this error is reduced to below 1$\%$ percent. Specially,
our calculation shows minimum error
 for energy 60-70 KeV  up to at N$^2$LO. Comparison of the LO
with the NLO and N$^2$LO results demonstrate convergence of the
effective field theory. Finally, three-body forces will enter at
higher orders of the EFT approach and reduce the theoretical
uncertainty.

%%%%%%%%%%%%%%%%%%%%%%%%%%%%%%%%%%%%%%%%%%%%%%%%%%%%%%%%%%%%%%%%%%%%%%%%%%%%%%%
\section{Acknowledgments}
%%%%%%%%%%%%%%%%%%%%%%%%%%%%%%%%%%%%%%%%%%%%%%%%%%%%%%%%%%%%%%%%%%%%%%%%%%%%%%%
The authors would like to thanks U. van Kolck for helpful
discussions. We would like to thanks  P.F. Bedaque and Harald W.
Grie\ss hammer for useful comments and valuable Mathematica code.
%%%%%%%%%%%%%%%%%%%%%%%%%%%%%%%%%%%%%%%%%%%%%

%%%%%%%%%%%%%%%%%%%%%%%%%%%%%%%%%%%%%%%%%%%%%%%%%%%%%%%%%%%%%
\bibliographystyle{apsrev}

\end{document}